\documentstyle[twoside,epsfig]{article}

\catcode`\@=11
\long\def\@makefntext#1{
\protect\noindent \hbox to 3.2pt {\hskip-.9pt  
$^{{\eightrm\@thefnmark}}$\hfil}#1\hfill}		

\def\@makefnmark{\hbox to 0pt{$^{\@thefnmark}$\hss}}	
	
\def\ps@myheadings{\let\@mkboth\@gobbletwo
\def\@oddhead{\hbox{}
\rightmark\hfil\eightrm\thepage}   
\def\@oddfoot{}\def\@evenhead{\eightrm\thepage\hfil
\leftmark\hbox{}}\def\@evenfoot{}
\def\sectionmark##1{}\def\subsectionmark##1{}}

\oddsidemargin=\evensidemargin
\newcounter{sectionc}\newcounter{subsectionc}\newcounter{subsubsectionc}
\renewcommand{\section}[1] {\vspace{12pt}\addtocounter{sectionc}{1} 
\setcounter{subsectionc}{0}\setcounter{subsubsectionc}{0}\noindent 
	{\tenbf\thesectionc. #1}\par\vspace{5pt}}
\renewcommand{\subsection}[1] {\vspace{12pt}\addtocounter{subsectionc}{1} 
	\setcounter{subsubsectionc}{0}\noindent 
	{\bf\thesectionc.\thesubsectionc. {\kern1pt \bfit #1}}\par\vspace{5pt}}
\renewcommand{\subsubsection}[1] {\vspace{12pt}\addtocounter{subsubsectionc}{1}
	\noindent{\tenrm\thesectionc.\thesubsectionc.\thesubsubsectionc.
	{\kern1pt \tenit #1}}\par\vspace{5pt}}

\newcounter{appendixc}
\newcounter{subappendixc}[appendixc]
\newcounter{subsubappendixc}[subappendixc]
\renewcommand{\thesubappendixc}{\Alph{appendixc}.\arabic{subappendixc}}
\renewcommand{\thesubsubappendixc}
	{\Alph{appendixc}.\arabic{subappendixc}.\arabic{subsubappendixc}}

\renewcommand{\appendix}[1] {\vspace{12pt}
        \refstepcounter{appendixc}
        \setcounter{figure}{0}
        \setcounter{table}{0}
        \setcounter{lemma}{0}
        \setcounter{theorem}{0}
        \setcounter{corollary}{0}
        \setcounter{definition}{0}
        \setcounter{equation}{0}
        \renewcommand{\thefigure}{\Alph{appendixc}.\arabic{figure}}
        \renewcommand{\thetable}{\Alph{appendixc}.\arabic{table}}
        \renewcommand{\theappendixc}{\Alph{appendixc}}
        \renewcommand{\thelemma}{\Alph{appendixc}.\arabic{lemma}}
        \renewcommand{\thetheorem}{\Alph{appendixc}.\arabic{theorem}}
        \renewcommand{\thedefinition}{\Alph{appendixc}.\arabic{definition}}
        \renewcommand{\thecorollary}{\Alph{appendixc}.\arabic{corollary}}
        \renewcommand{\theequation}{\Alph{appendixc}.\arabic{equation}}
        \noindent{\tenbf Appendix#1}\par\vspace{5pt}}
\newcommand{\subappendix}[1] {\vspace{12pt}
        \refstepcounter{subappendixc}
        \noindent{\bf Appendix \thesubappendixc. {\kern1pt \bfit #1}}
	\par\vspace{5pt}}
\newcommand{\subsubappendix}[1] {\vspace{12pt}
        \refstepcounter{subsubappendixc}
        \noindent{\rm Appendix \thesubsubappendixc. {\kern1pt \tenit #1}}
	\par\vspace{5pt}}

\newcommand{\textlineskip}{\baselineskip=13pt}
\newcommand{\smalllineskip}{\baselineskip=10pt}

\def\eightcirc{
\begin{picture}(0,0)
\put(4.4,1.8){\circle{6.5}}
\end{picture}}
\def\eightcopyright{\eightcirc\kern2.7pt\hbox{\eightrm c}} 

\newcommand{\copyrightheading}[1]
	{\vspace*{-2.5cm}\smalllineskip{\flushleft
	{\footnotesize International Journal of Theoretical and Applied Finance#1}\\
	{\footnotesize $\eightcopyright$\, World Scientific Publishing
	 Company}\\
	 }}

\newcommand{\pub}[1]{{\begin{center}\footnotesize\smalllineskip 
	#1\\		
	\end{center}
	}}

\def\abstracts#1#2#3{{
	\centering{\begin{minipage}{4.5in}\baselineskip=10pt\footnotesize
	\parindent=0pt #1\par 
	\parindent=15pt #2\par
	\parindent=15pt #3
	\end{minipage}}\par}} 

\renewenvironment{thebibliography}[1]
	{\frenchspacing
	 \ninerm\baselineskip=11pt
	 \begin{list}{\arabic{enumi}.}
        {\usecounter{enumi}\setlength{\parsep}{0pt}     
	 \setlength{\leftmargin 12.7pt}{\rightmargin 0pt} 
         \setlength{\itemsep}{0pt} \settowidth
	{\labelwidth}{#1.}\sloppy}}{\end{list}}

\newcounter{itemlistc}
\newcounter{romanlistc}
\newcounter{alphlistc}
\newcounter{arabiclistc}

\newcommand{\fcaption}[1]{
        \refstepcounter{figure}
        \setbox\@tempboxa = \hbox{\footnotesize Fig.~\thefigure. #1}
        \ifdim \wd\@tempboxa > 5in
           {\begin{center}
        \parbox{5in}{\footnotesize\smalllineskip Fig.~\thefigure. #1}
            \end{center}}
        \else
             {\begin{center}
             {\footnotesize Fig.~\thefigure. #1}
              \end{center}}
        \fi}

\newcommand{\tcaption}[1]{
        \refstepcounter{table}
        \setbox\@tempboxa = \hbox{\footnotesize Table~\thetable. #1}
        \ifdim \wd\@tempboxa > 5in
           {\begin{center}
        \parbox{5in}{\footnotesize\smalllineskip Table~\thetable. #1}
            \end{center}}
        \else
             {\begin{center}
             {\footnotesize Table~\thetable. #1}
              \end{center}}
        \fi}

\def\@citex[#1]#2{\if@filesw\immediate\write\@auxout
	{\string\citation{#2}}\fi
\def\@citea{}\@cite{\@for\@citeb:=#2\do
	{\@citea\def\@citea{,}\@ifundefined
	{b@\@citeb}{{\bf ?}\@warning
	{Citation `\@citeb' on page \thepage \space undefined}}
	{\csname b@\@citeb\endcsname}}}{#1}}

\newif\if@cghi
\def\cite{\@cghitrue\@ifnextchar [{\@tempswatrue
	\@citex}{\@tempswafalse\@citex[]}}
\def\citelow{\@cghifalse\@ifnextchar [{\@tempswatrue
	\@citex}{\@tempswafalse\@citex[]}}
\def\@cite#1#2{{$\null^{#1}$\if@tempswa\typeout
	{IJCGA warning: optional citation argument 
	ignored: `#2'} \fi}}

\def\pmb#1{\setbox0=\hbox{#1}
	\kern-.025em\copy0\kern-\wd0
	\kern.05em\copy0\kern-\wd0
	\kern-.025em\raise.0433em\box0}

\def\fnt#1#2{\footnotetext{\kern-.3em
	{$^{\mbox{\scriptsize #1}}$}{#2}}}

\def\fpage#1{\begingroup
\voffset=.3in
\thispagestyle{empty}\begin{table}[b]\centerline{\footnotesize #1}
	\end{table}\endgroup}

\def\runninghead#1#2{\pagestyle{myheadings}
\markboth{{\protect\footnotesize\it{\quad #1}}\hfill}
{\hfill{\protect\footnotesize\it{#2\quad}}}}
\font\tenrm=cmr10
\font\tenit=cmti10 
\font\tenbf=cmbx10
\font\bfit=cmbxti10 at 10pt
\font\ninerm=cmr9

\font\eightrm=cmr8

\def\qed{\hbox{${\vcenter{\vbox{			
   \hrule height 0.4pt\hbox{\vrule width 0.4pt height 6pt
   \kern5pt\vrule width 0.4pt}\hrule height 0.4pt}}}$}}

\def\theequation{\thesectionc.\arabic{equation}}	

\begin{document}

\runninghead{Optimal lag in dynamical investments}
{Optimal lag in dynamical investments}

\normalsize\textlineskip
\thispagestyle{empty}
\setcounter{page}{1}

\copyrightheading{}			
\vspace*{0.88truein}

\fpage{1}
\centerline{\bf OPTIMAL LAG IN DYNAMICAL INVESTMENTS}
\vspace*{0.37truein}

\centerline{\footnotesize MAURIZIO SERVA}
\vspace*{0.015truein}
\centerline{\footnotesize\it Dipartimento di Matematica, Universit\`a dell'Aquila, 
and I.N.F.M.}
\baselineskip=10pt
\centerline{\footnotesize\it I-67010 Coppito, L'Aquila, Italy}

\vspace*{0.225truein}
\pub{ \today }

\vspace*{0.21truein}
\abstracts{
A portfolio of different stocks and a risk-less security
whose composition is dynamically maintained stable
by trading shares at any time step leads to a 
growth of the capital with a nonrandom rate. 
This is the key for the theory of optimal-growth investment 
formulated by Kelly. In presence of transaction costs,
the optimal composition changes and, more important,
it turns out that the frequency of transactions must be reduced. 
This simple observation leads to the definition of an optimal lag
between two rearrangement of the portfolio.
This idea is tested 
against an investment in a risky asset and a risk-less one.
The price of the first is proportional to NYSE composite index
while the price of the second grows according to the 
American Discount Rate.
An application to a portfolio of many stochastically equivalent securities
is also provided.
}{}{}

\vspace*{1pt}\textlineskip	
\section{Introduction}	
\vspace*{-0.5pt}
\noindent

The definition of an optimal portfolio is a challenging problem in 
theoretical finance~\cite{BouchaudPotters,Merton,Ingersoll,Litner,Markowitz} 
and it has an obvious relevance in technical studies.
Suggestions and indications for investments can be found 
in any economic newspaper where, usually, reference is to static strategies.
The problem, in this case, consists in finding the
best initial composition according to 
the risk attitudes of the investor
and later trading is not expected.
Nevertheless, if transaction costs are negligible,
it turns out that it is rentable to maintain stable
the composition of the portfolio by selling or buying shares.
According to this dynamical point of view, the fraction of the capital
invested in any stocks or security may be kept constant in time.
The most important consequence
is that the investor wealth grows with a nonrandom rate
when the investment is repeated many times.
This fact, which is a trivial consequence of
the law of large numbers, implies
that the optimal-grow strategy is the only possible,
while subjective risk averseness or other psychological 
considerations play no role.

This point is still often misunderstood in the current literature.
For example, Samuelson and Merton~\cite{Samuelson,MertonSamuelson} 
demonstrated that the growth-optimal strategy does not
maximize the expected value of a generic utility function.
Nevertheless, an investor
which would decide to optimize her strategy with respect to a generic
utility function would, {\it almost surely},
end up with an exponentially smaller capital.
The reason is that the dominant contribution to the expected value comes from
events whose probability exponentially vanishes in time.
This is a general probabilistic fact,
widely studied in the context of large deviations theory.
We should stress once again that the above considerations 
applies whenever one deals with long time repetition of the same investment.
On the contrary, they do not apply to strategies concerning
static investments, as for example 
the composition of a portfolio of securities which remains unchanged
until they expire or they are sold out.

In this paper we extend Kelly theory showing that 
it still holds when transaction costs are considered.
Nevertheless, in this case, it is better to reduce the frequency of trading.
This simple observation leads to the definition of an optimal 
lag between transactions.
In section 2 we summarize Kelly theory assuming that
interest rate may vary in time. In section 3
we analyze the effects due to
transaction costs which are assumed to be proportional to the
amount of shares traded and we show how the 
notion of an optimal lag naturally emerges.
In section 4 we test our result 
against a portfolio with a risky asset and a risk-less one.
We first consider a realistic situation
where the price of the first is proportional to NYSE composite index
while the price of the second grows according to the 
American Discount Rate. Then, for the sake of comparison, 
we reconsider the classical coin toss game originally proposed by Kelly.
An application to a portfolio of many stochastically 
equivalent securities is provided in section 5. We first show that,
in absence of transaction cost, high frequency trading
allows for a positive growth rate of the capital even 
when a static investment leads to a vanishing rate.
Than, we show that advantageous dynamical investment
is still possible in presence of trading 
costs. In this case the optimal lag scales non trivially
with costs. Finally, in section 6, we shortly discuss the 
relevance of our results with respect to 
the notion of continuous time in finance. 

\vspace*{1pt}\textlineskip	
\section{The Kelly theory of Optimal Gambling}
\vspace*{-0.5pt}
\noindent

The theory of optimal-growth investment was formulated by
Kelly~\cite{Kelly} in a contest not directly related to finance and
stock market. His original purpose was mainly to
find an interpretation of the Shannon~\cite{Shannon} entropy
in terms of optimal gambling strategies.
This theory was later reconsidered in a more finance related contest 
by Breiman~\cite{Breiman1960,BreimanBerkeley},
more recently it has been rediscovered and extended by various 
authors~\cite{GalluccioZhang,MaslovZhang,MarsiliMaslovZhang,BavieraPasquiniServaVulpiani,Slanina}
and it has been also applied to the problem of pricing 
derivatives~\cite{AurellBavieraHammarlidServaVulpiani,AurellBavieraHammarlidServaVulpiani2} in the general case 
of incomplete markets.

Consider a stock, or some other security, whose price is described by
\begin{equation} 
S_{t+1}=u_t S_{t}
\label{udef}
\end{equation} 
where time is discrete, $S_t$ is the price at time $t$ 
of a share and the $u_t$ are independent, identically distributed
random variables.
Also assume that the risk-less interest rate $r_t$ may vary in time.

Consider now an investor who starts at time $0$ with a
wealth $W_0$, and who decides to invest in this stock many times.
Suppose that she chooses to invest at each time
a fraction $l$ of her capital in stock,
and the remaining part in a risk-less security, i.e. a bank account
with rate $r_t$.
In absence of transaction costs,
her wealth evolves as a multiplicative random process
\begin{equation} 
W_{t+1}= (1-l)r_t W_t +l u_t W_t \,\, .
\label{Wdef1}
\end{equation}
It is useful to introduce the discounted prices
$\tilde{u}_t \equiv u_t/r_t$, so that (\ref{Wdef1})
rewrites as
\begin{equation} 
W_{t+1}=r_t\left(1+l(\tilde{u}_t-1)\right)W_t\,\, .
\label{Wdef2}
\end{equation}
In the large time limit we have, by the law of large numbers,
that the exponential growth rate of the wealth is, 
with probability one, a constant.
That is,
\begin{equation}
\lambda(l) \equiv \lim_{T\to\infty}\frac{1}{T} 
\log\frac{W_T}{W_0}\,\, .
\label{Gdef}
\end{equation}
It is clear from (\ref{Wdef1}) that the interest $r_t$
contributes to the above limit with an additive term 
which is independent from the strategy, and corresponds to the rate 
which the investor would obtain by investing all her capital 
in the risk-less security (bank).
Therefore, the general problem
with a time dependent $r_t$, can be always
mapped into the $r_t =1$ problem by properly discounting
the security prices.
In the second part of this section
we drop the tildes, assuming that all prices are already discounted.
The net growth rate (the relative growth rate with respect to
a capital entirely invested in the risk-less security)
is than, for almost all realizations of the
random variables $u_t$,
\begin{equation}
\lambda(l)= E[\log\left(1+l(u-1)\right)]
\label{lambdef}
\end{equation}
where $E[\, \cdot \,]$
represent the average with respect to the distribution of the $u$.

The optimal gambling strategy of Kelly consists in maximizing
$\lambda(l)$ with respect to $l$.
The solution is unique because the logarithm is a convex function of
its argument:
\begin{equation}
\lambda^*=\max_{l} \lambda(l)= \lambda(l^*)\,\, .
\label{lmax}
\end{equation}

Notice that an investor can never have a negative capital,
which implies that $1+l(u-1)$ must be always positive.
This is same to say that the argument of the
logarithm must be positive.
Therefore one must have that 
$l<1/(1-u_{min})\equiv l_{max}$
where $u_{min}$ is the minimum value that
the stochastic variable $u$ can assume.
Also notice that, at variance with the original formulation of Kelly,
the investor is allowed to borrow money, so that
$l$ can also take values larger than the unity (but lower than $l_{max}$).
Only when $u_{min} =0$ the investor is not
allowed to borrow money.

\vspace*{1pt}\textlineskip	
\section{Transaction costs and optimal lag.}	
\vspace*{-0.5pt}
\noindent
In this section we consider the effects due to transaction 
costs.
This problem, which is a classical topic in mathematical finance
(see for example~\cite{AkianMenaldiSulem,MagillConstantinides,ShreveSoner,Zariphopoulou}),
is here reconsidered with the aim of 
defining an optimal lag for transactions.

Suppose that at time $t$ the agent
invest a part $l W_t$ of her capital in the stock,
after a  time step, the capital in the stock has 
become $l u_{t} W_t$.
Then she wants to restore the previous proportion, 
so that the capital invested in the stock is $l W_{t+1}$.
In this case, she has to sell or buy the exceeding or missing shares.
The entire process, assuming a trade cost 
proportional to the value of the traded shares,
is described by the implicit equation
\begin{equation}
W_{t+1} = (l u_{t} +1-l)W_t -\gamma
|l u_{t}W_t - l W_{t+1}|
\label{Wgamma}
\end{equation}
where $\gamma$ is the proportionality constant (see also~\cite{Slanina}).
This equation can be made explicit and one obtains
\begin{equation}
W_{t+1} = A(u_t , l , \gamma) W_t
\label{Wgammaexp}
\end{equation}
where
\begin{equation}
A(u , l , \gamma)=
\frac{1+l(u-1)+\alpha \gamma l u}{1+\alpha \gamma l}
\label{afactor}
\end{equation}
and
\begin{equation}
\alpha \equiv sign(u-1) \, sign(l-1)\,\, .
\label{alpha}
\end{equation}

Notice that according to this simple rule,
the behaviour of an investor is qualitatively different
when $l$ is larger or smaller than 1.
In the first case, in fact, one has a speculative behaviour:
some of the shares are sold out after their price has decreased.
In the second case, one has a prudent behaviour:
some of the shares are sold out when their prise has increased.

The resulting rate (for a given $\gamma$ and
for a given probability 
for the $u$) will be a function of $l$ and will depend on
$\gamma$
\begin{equation}
\lambda_\gamma(l) =
E[\log A(u, l , \gamma)]\,\, .
\label{lgamma}
\end{equation}
The optimal rate will be chosen by finding the fraction
$l$ which maximizes the above expression.

For increasing value of $\gamma$ the amount of transactions must become
smaller and the optimal $l$ has to approach one of the two limits 
which corresponds to a fixed portfolio: $l=1$ or $l=0$.
The choice between the two depends on the distribution of the $u$,
if $E[log(u)]>0$, than all the capital will be in the stock 
($l=1 , \lambda_\gamma=E[log(u)]$),
otherwise, all the capital will be in the risk less security 
($l=0, \lambda_\gamma=0$).

It is clear, at this point, 
that in presence of trading costs,
it would be convenient to rearrange the capital
less frequently. In other words, 
between the two limiting strategies,
the static and the extremely dynamical one,
it is possible to find a compromise.
One can decide to rearrange the composition of the portfolio only 
every $\tau$ time steps.
This strategy only leads to a redefinition of the 
reference stochastic variable.
In fact, once defined 
\begin{equation}
U_{t,\tau} \equiv \prod_{i=t+1}^{t+\tau} u_{i}
\label{U}
\end{equation}
one ends up with the evolution law
\begin{equation}
W_{t+\tau} = A(U_{t,\tau} , l , \gamma) W_t
\label{eet}
\end{equation}
where $A$ has the same form as before.
The associated rate of growth of the capital is
\begin{equation}
\lambda_\gamma (\tau , l)=
\frac{1}{\tau} E[\log A(U_\tau, l , \gamma)]
\label{l}
\end{equation}
where $t$ has been eliminated from the notation
because of the time translation invariance.
The rate has to be maximized both with respect to
$\tau$ and $l$.
\begin{equation}
\lambda_\gamma^* = \max_{l,\tau}\lambda_\gamma (l,\tau) =
 \max_{\tau}\lambda_\gamma (l(\tau),\tau) = \lambda_\gamma (l^*,\tau^*)\,\, .
\end{equation}
Notice that, in absence of transaction costs, 
the optimal lag $\tau^*$ is always the minimal one ($\tau^*=1$). 
On the contrary, it may happen that, for large transaction costs,
$\tau^*$ becomes infinite, i.e.
the static strategy turns out to be the best.

\vspace*{1pt}\textlineskip	
\section{Real example from NYSE index}	
\vspace*{-0.5pt}
\noindent

In order to show how this idea works in practice we consider
a security whose price is proportional to the NYSE composite index.
We will look to its price movement for exactly one decade,
from the $1^{st}$ of September 1988 to the $31^{th}$ of August 1998.
First of all we have to give an estimation of the the risk-less
interest rate $r_t$. The simplest thing to do is
to look at the American Discount Rate $R_t$ during the same period
which is plotted in fig. 1 (in $\%$).

\begin{figure}
\vspace*{13pt}
\begin{center}
\mbox{\epsfig{file=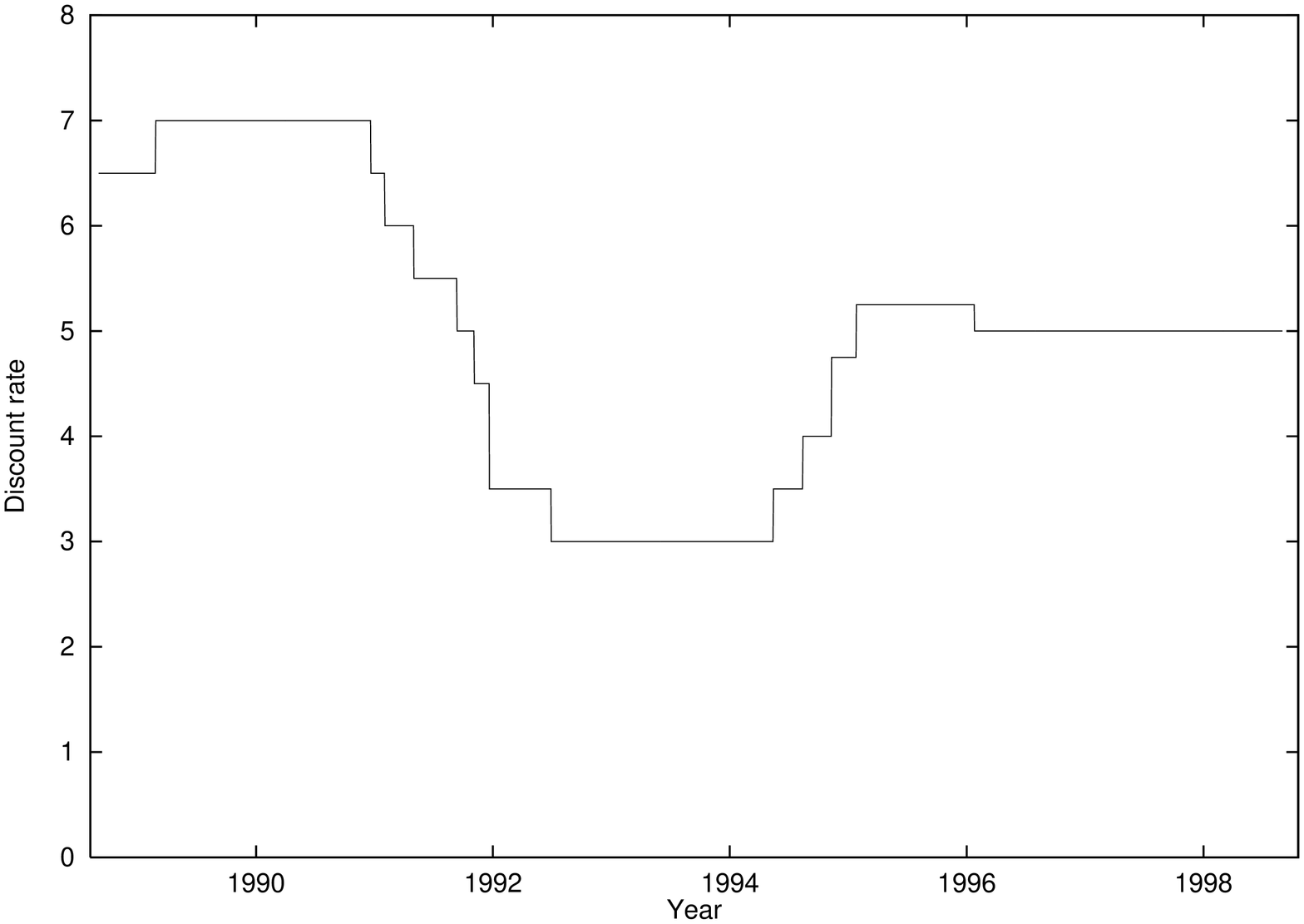,width=3.5in,angle=270}}
\end{center}
\vspace*{13pt}
\fcaption{
American Discount Rate in $\%$ from the $1^{st}$ of September 1988 to the 
$31^{th}$ of August 1998.
}
\label{dx}
\end{figure}

Far from being a constant , $R_t$ ranges  from 3 to 7.
Then a good estimation of $r_t$ is 
\begin{equation}
r_t = (1+\frac{R_t}{100})^{\frac{1}{253}}\,\, .
\end{equation}
It is then easy to obtain from the NYSE index $S_t$
its discounted counterpart
\begin{equation}
\tilde{S}_t =\frac{S_t}{\prod_{i=0}^{t-1} r_i}\,\, .
\end{equation}
In fig. 2  we plot the discounted NYSE index
whose initial value has been put equal the unity.
Notice that a capital entirely invested in the stock
would double in ten years with respect to the same 
capital invested in the risk-less security.

\begin{figure}
\vspace*{13pt}
\begin{center}
\mbox{\epsfig{file=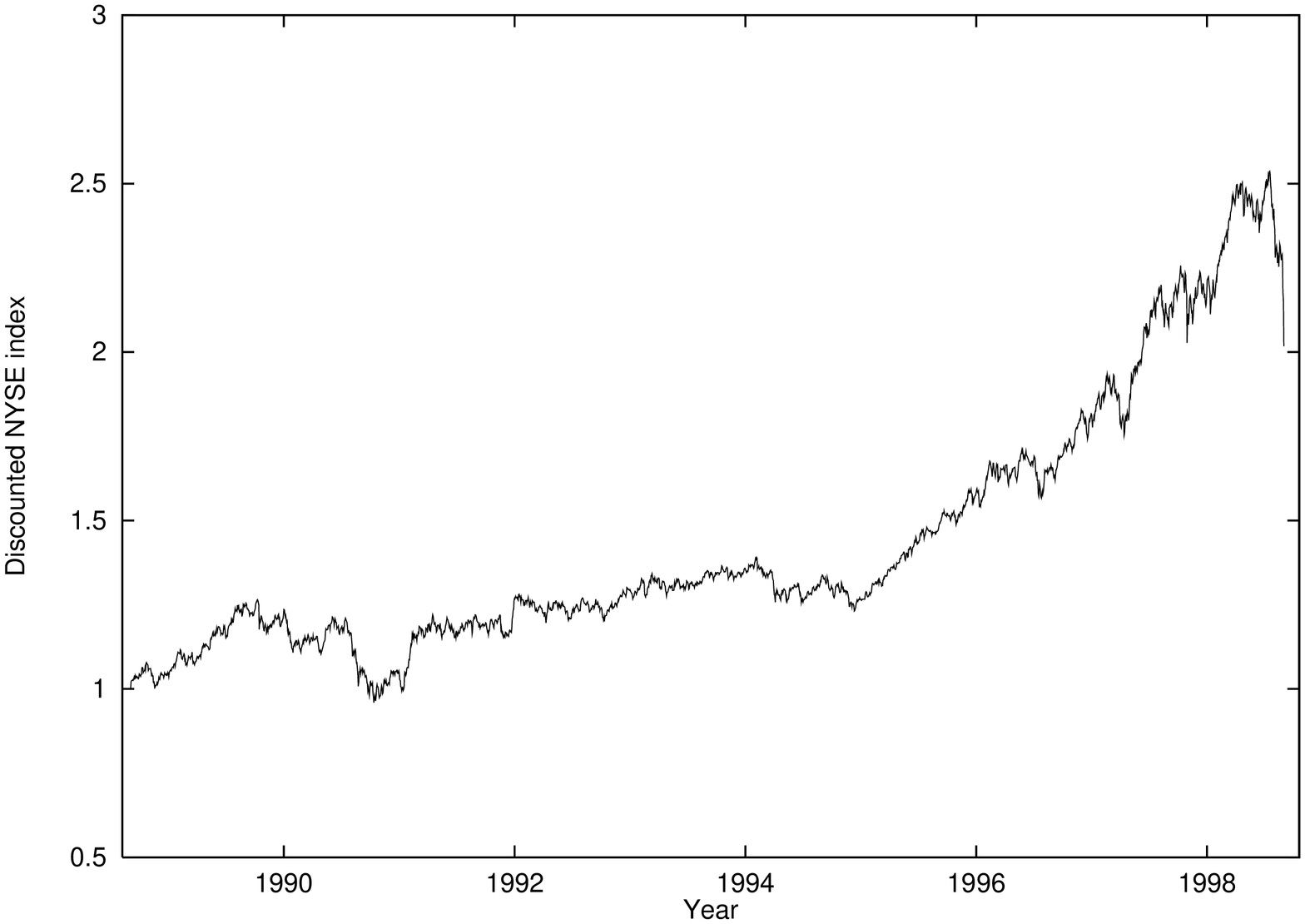,width=3.5in,angle=270}}
\end{center}
\vspace*{13pt}
\fcaption{
Nyse discounted index from the $1^{st}$ of September 1988 
to the $31^{th}$ of August 1998.
}
\label{nydx}
\end{figure}

Using this data, we disregard all correlations,
and we assume that all increments are independent. 
Than it is easy to compute the
final value of the capital for different choices
of $l$ assuming that its initial value is 1
and that $\gamma=0$.

In fig. 3 we plot the final capital
$\exp\{\lambda(l) T\}$ as a function of $l$
for vanishing transaction costs and 
for three different values of $\tau$ corresponding to one day, 
one week (5 working days) and one month (21 working days).
Obviously, the best time lag will 
be the minimal one ($\tau=1$, full line),
in this case the maximum is reached for $l\simeq 5$,
implying that the optimal investment in stock
should be five time larger than the owned capital.
The maximum corresponds to a final capital which
is about about seven times the initial capital, 
much larger than the static result ($l=1$) which gives a
final capita only twice larger than the initial one. 
For $l$ larger than 14 the capital vanishes,
which implies that $u_{min}=0.93$. 
The dashed and the dotted lines correspond to
the more static strategies of rearranging 
the portfolio every week and every month. 
We clearly see that, as expected, these more static strategies are less 
efficient than the dynamical one.
\begin{figure}
\vspace*{13pt}
\begin{center}
\mbox{\epsfig{file=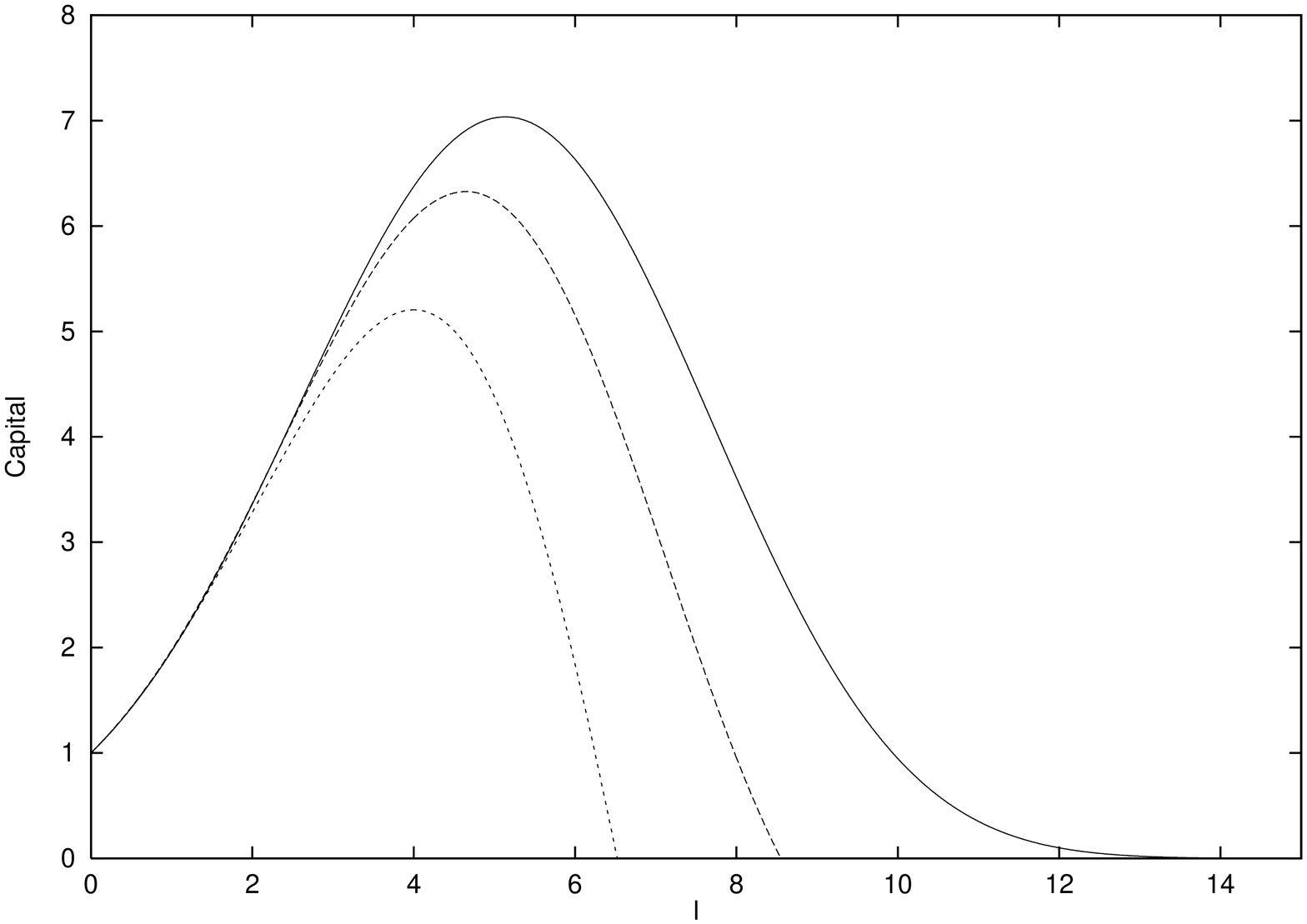,width=3.5in,angle=270}}
\end{center}
\vspace*{13pt}
\fcaption{
The final capital $\exp(\lambda^*(l)T)$ versus $l$
for $\gamma=0$ and for three different values of $\tau$ 
(the full line is one day, 
the dashed line is one month, the dotted line is one year).
}
\label{nyopl}
\end{figure}

In fig. 4 we consider exactly the same situation
for $\gamma =0.003$.
In this case the best result corresponds
to arrangements every week (dashed line).
The optimal $l$ is about 4, smaller than the cost-free result,
and the final capital is now only five time larger that the initial one.
The daily strategy (full line) is much less efficient
for this value of $\gamma$, while there is a very small
difference with the more static strategy corresponding
to monthly transactions.

It may be useful to test the general strategy 
against the classical Kelly coin toss. 
In this game one has
that $u= 2$ with probability $p$ and $u = 0$ with probability $1-p$.
If a lag $\tau$ and a fraction $l$ are chosen,
than the corresponding growth rate is
\begin{eqnarray}
\lambda_\gamma (\tau,l)=
\frac{p^\tau}{\tau}
\log\left( \frac{1+l(2^\tau-1) -2^\tau 
\gamma l }{ 1-\gamma l }\right) \nonumber \\
+\frac{1-p^\tau}{\tau}
\log \left(\frac{1-l}{1+\gamma l}\right)\,\, .
\label{lgammak}
\end{eqnarray}

\begin{figure}
\vspace*{13pt}
\begin{center}
\mbox{\epsfig{file=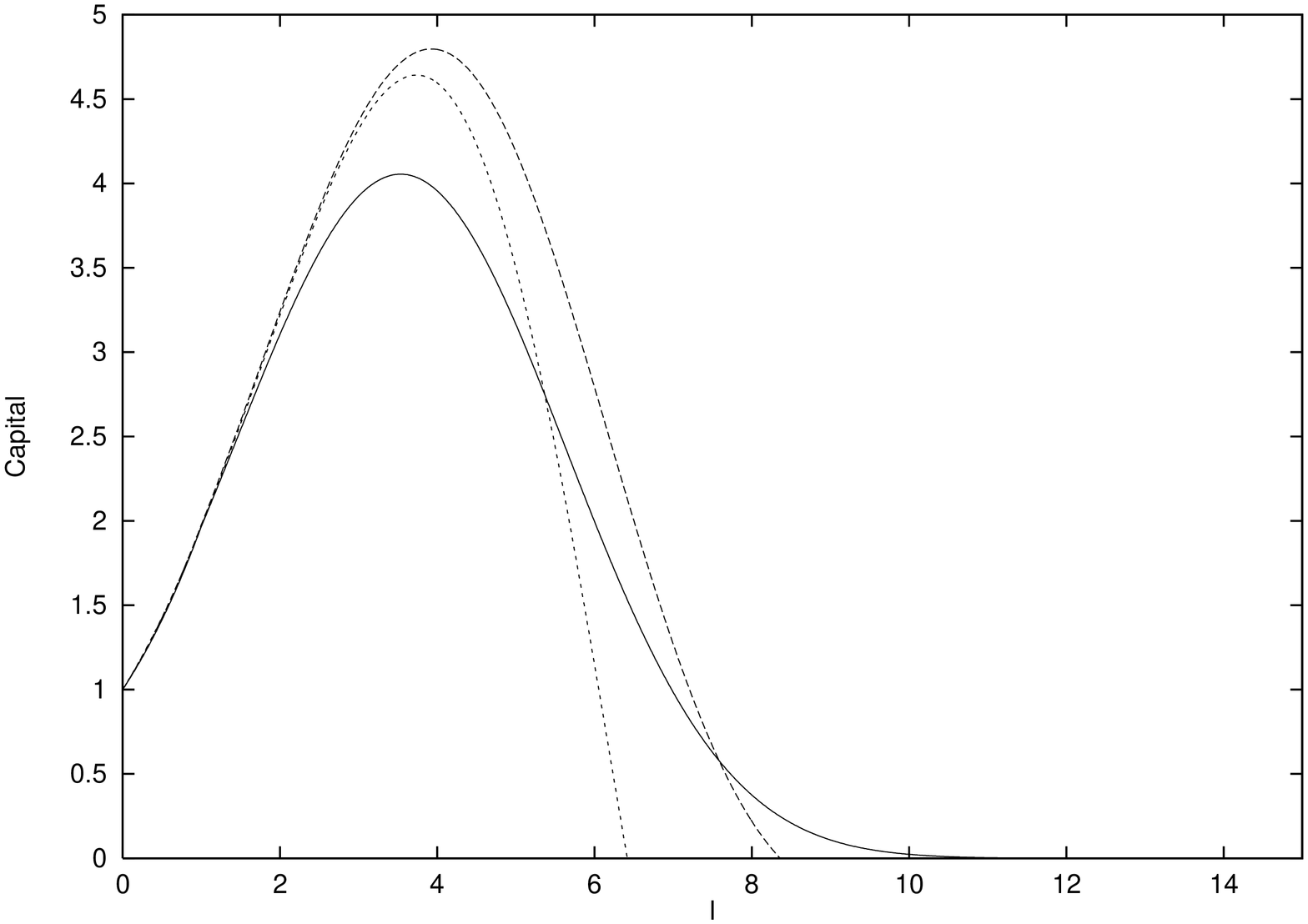,width=3.5in,angle=270}}
\end{center}
\vspace*{13pt}
\fcaption{
The final capital $\exp(\lambda^*(l)T)$ versus $l$
for $\gamma=0.003$ and for three different values of $\tau$ 
(the full line is one day, 
the dashed line is one month, the dotted line is one year).
}
\label{nytaugamma}
\end{figure}

A consequence of the above formula is that 
the minimum probability $p$ necessary to
have a positive rate  when $\tau=1$ is 
\begin{equation}
p_{min} = \left( \frac{1+\gamma}{2} \right)
\label{pmin}
\end{equation}
which says that for $p<p_{min}$ it is better 
to do not invest at all in the stock, if lags 
larger than 1 are not allowed.

If the rate (\ref{lgammak}) is maximized with respect to $l$
one obtains $\lambda_\gamma (l(\tau),\tau)$.
It is useful to plot this quantity (for given $\gamma$ and $p$)
with respect to $\tau$, in order to compare with the
classical Kelly result, and in order to have a qualitative idea
on the conditions for a non trivial optimal $\tau^*$.
In fig. 5 we plot the relative rate
$ \lambda_\gamma (l(\tau),\tau)/ \lambda^*$ versus $\tau$,
where $\lambda^*=\log(2) +p\log p +(1-p)\log(1-p)$ 
is the cost-less Kelly optimal rate.
In absence of costs (full line) we have that 
the relative rate equals 1 at $\tau=1$, i.e. we recover the Kelly 
result. The line, as expected, monotonically decreases for larger lags,
and vanishes for lags of about 10.
The same qualitative behaviour also is found for 
$\gamma=0.0005$ (dashed line) and $\gamma=0.001$ (dotted line),
the best lag being still the minimal one.
The only difference is that now the cost-less
Kelly rate cannot be entirely recovered.
Only for $\gamma=0.002$ the qualitative behaviour changes and the optimal 
$\tau^*$ turns out to be about 4. 

\begin{figure}
\vspace*{13pt}
\begin{center}
\mbox{\epsfig{file=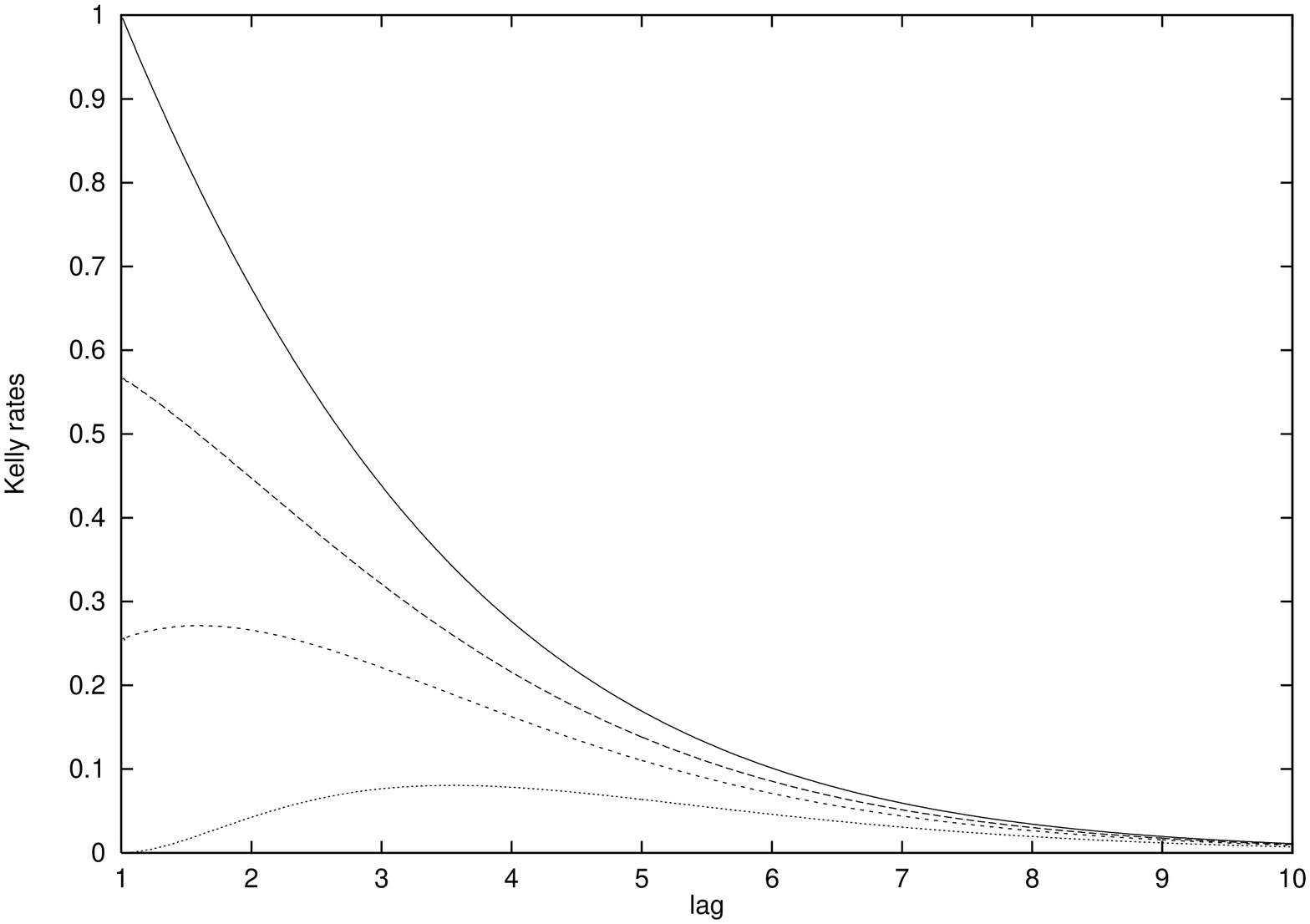,width=3.5in,angle=270}}
\end{center}
\vspace*{13pt}
\fcaption{
Kelly dichotomic relative growth rate versus $\tau$
for $p=0.51$ and for different values of $\gamma \, 
(0,0.005,0.01,0.02)$.
}
\label{ktau}
\end{figure}

\vspace*{1pt}\textlineskip	
\section{Diversified portfolios}
\vspace*{-0.5pt}
\noindent

The portfolio problem we have considered in previous sections
only allows for two different choices: a stock and a 
risk-less security. It is useful to reconsider the problem
from the point of view of an agent which can chose 
to invest her capital in many different stocks.
For the sake of simplicity assume that the prices of all stocks evolve
independently but with the same probabilistic law.
Suppose the number of stocks is $N$ 
and that the price changes according to
\begin{equation}
S_{t+1}^{(k)} = u_{t}^{(k)}  S_{t}^{(k)} 
\label{spk}
\end{equation}
where $k=1,......,N$ and the $u_{t}^{(k)}$ are 
equally distributed and independent both in time
and for different stocks.
Also assume that
\begin{equation}
E[log(u)]=0   \,\,\,\,\, \,\,\,\, Var[log(u)]=\sigma^2\,\, .
\end{equation}
The first of these two assumptions, only means that we consider
de-meaned returns, the effect of
a positive constant trend being completely trivial.
To simplify notation let us write that
$u_{t}^{(k)} = exp(\eta_{t}^{(k)})$,
where the $\eta$ are independent variables,
with vanishing mean, and variance $\sigma^2$.

We will now consider first an investment strategy
in absence of transaction costs,
showing the advantage of a dynamical approach to 
the problem. We will then introduce transaction costs 
showing that in this case the optimal lag 
increases as a power of $\gamma$.
 
\subsection{Vanishing transaction costs}
\noindent

Let us consider the general strategy corresponding to
an arbitrary lag $\tau$ in absence of trading costs..
As in the single stock case, we define  
\begin{equation}
U^{(k)}_{t,\tau} \equiv \prod_{i=t+1}^{t+\tau} u^{(k)}_{i}\,\, .
\label{Uk}
\end{equation}
The new variables can be rewritten with the
previous notation,$U^{(k)}_{t,\tau} = exp(\eta_{t}^{(k)}\sqrt{\tau}$,
where the $\eta$ are, as before independent variables,
with vanishing mean, and variance $\sigma^2$.
Let us also define
\begin{equation}
\bar{U}_{t,\tau} \equiv
\frac{1}{N}\sum_{k=1}^N U_{t,\tau}^{(k)}\,\, .
\end{equation}

Because of the symmetry of the problem we 
can safely assume that the agent rearrange her capital
in order to have a fraction $W_t/N$ invested
in each stock at the beginning of any period of length $\tau$.
In this case, the capital would grow with a rate
\begin{equation}
\lambda(\tau) = \frac{1}{\tau}
E[ \log \bar{U}_{\tau} ] \,\, .
\label{lnf}
\end{equation}

Then, assuming that $\sigma^2 \tau \ll 1$, 
one has the approximate result (Taylor expansion
up to order four in $\sigma^2 \tau$)
\begin{equation}
\lambda(\tau) \simeq \frac{N-1}{N} \frac{\sigma^2}{2}
 -\frac{\sigma^4 \tau}{4N}\,\, .
\label{lnf2}
\end{equation}
Notice that, as expected, it is convenient to rearrange the
capital every day, since $\lambda(\tau)$
is monotonically decreasing.
In this way, in fact, the investor is able to take advantage
from the fact that she is investing in many stocks,
ant the rate depends on the volatility of the prices.
For very large values of $\tau$, the above expansion 
does not hold, nevertheless, one has $\lambda(\tau)\to 0$
when $\tau \to \infty$.
This fact implies that a static investor, 
at variance with the dynamical
one, is not able to take advantage from 
the portfolio diversification.
To resume, the dynamical strategy $\tau=1$ allows for 
a positive rate close to $\sigma^2 /2$ at variance with
the static one $\tau=\infty$ which gives a vanishing rate.

\subsection{Non vanishing transaction costs}
\noindent

Assume as before that the agent invests a fraction 
$W_t/N$ of her capital in any of the $N$ stocks
at the beginning of any period of length $\tau$.
Then she waits till time $t+\tau$. At this later time the money in 
the stock $k$ will be
$U_{t,\tau}^{(k)} W_t/N$.
It is clear that the capital is not anymore 
equally distributed between the $N$ stocks. 
An equal distribution of the capital would give an amount
$ W_{t+\tau}/N $ in any stock.
The difference between the two amounts represents
the quantity of shares of the $k$ stock
to sell or to buy in order to reconstruct a portfolio were 
the capital is equally distributed between stocks.
Then the cost
of the operation of buying or selling 
shares of the $k$ stock will be
\begin{equation}
\frac{\gamma}{N} \, | U_{t,\tau}^{(k)}W_t - W_{t+\tau}| 
\simeq
\frac{\gamma}{N} \, | U_{t,\tau}^{(k)} - \bar{U}_{t,\tau}|W_t
\label{tck}
\end{equation}
where the approximation holds
up to terms of the second order in $\gamma$.
The total operation of redistribution will cost the sum over $k$ of
the above single stock cost.

Due to this redistribution,
the typical rate of grow of the capital will change in
\begin{equation}
\lambda_\gamma (\tau) = \frac{1}{\tau}
E[ \log \left( \bar{U}_{\tau} - \frac{\gamma}{N}
\sum_{k=1}^N  |U_{\tau}^{(k)}- \bar{U}_{\tau}| \right) ] 
\label{ln}
\end{equation}
where $t$ has been eliminated from the notation because of
the time translation invariance.

The problem is to find the $\tau^*$ which optimize the rate.
For vanishing transaction costs we have seen that
$\tau^*$ is $1$, while we now expect larger values of $\tau^*$. 
In order to estimate $\tau^*$ we assume again 
$ \sigma^2 \tau \ll 1$. As before we expand 
up to the fourth order in $\sigma^2 \tau$ and 
now we also expand up to the first order in $\gamma$.
Assuming that $E[|\eta|]= c \sigma$ ($c\le 1$), 
one obtains

\begin{equation}
\lambda_\gamma (\tau) \simeq
(1-\frac{1}{N}) \frac{\sigma^2}{2} -
\frac{\sigma^4 \tau}{4N} 
-\frac{c\gamma\sigma}{\tau^\frac{1}{2} }  
\label{lap}
\end{equation}
where terms of the order of $\gamma/N$ have been neglected.
The above expression as a maximum for

\begin{equation}
\tau^* =
\frac{(2c\gamma N)^{\frac{2}{3}}}{\sigma^{2}}\,\, .
\label{ts}
\end{equation}

The optimal time is, therefore,
proportional to $\gamma^{\frac{2}{3}}$,
until the expansion remains valid.
This $\tau^*$ corresponds to a rate 
\begin{equation}
\lambda_\gamma^* \simeq
\frac{N-1}{N} \frac{\sigma^2}{2}
-\frac{3}{2} 
\frac{(c\gamma)^\frac{2}{3}}{(2N)^\frac{1}{3} } \sigma^2\,\, . 
\label{16}
\end{equation}

This result also assures that a dynamical,
strategy allows for a positive rate also in presence of transaction costs 
provided that the expansion is 
self-consistent ($\sigma^{*2}\tau \ll 1$) and 
the rate $\lambda_\gamma^*$ remains positive.
This inequalities are verified when $\gamma N <1$.
For example a realistic evaluation for shares is $\gamma=0.01$,
which is compatible with a portfolio of less than one
hundred different stocks.

\vspace*{1pt}\textlineskip	
\section{Final Remarks}
\vspace*{-0.5pt}
\noindent
The proposal of this paper is to introduce the notion of 
an optimal lag for transactions in order to 
bridge between static and dynamical portfolio strategies.
The lag is chosen to be a deterministic quantity,
nevertheless, one could choose more refined strategies
in which it is allowed to be a stochastic variable.
For example, one could chose to or to sell
some of the shares when the
composition of the portfolio becomes
sufficiently far from the optimal one.
Such a strategy 
implies that lags depend on the 
evolution of the price and their probability distribution 
can be found out in the context of first hitting time theory.

Nevertheless, also in this case
the important fact is that lags are discrete.
The consequence is that the idea
of a continuous trading time turns out to be only a fictitious
assumption, even when an asset price is established 
with high frequence.
The lag between transactions, in fact, is usually much larger
than lag between two consecutive fixing of a price. 

This simple consideration has relevance 
for the classical problem of derivative pricing.
The most successful approach, due to Black and Scholes,
works for a complete market,
which means that trading time is 
assumed to be continuous.
In the light of the present discussion,
it is clear that a complete market only can be considered as an approximation
and more realistic pricing, accounting for incomplete markets
(i.e. discrete lags), 
has to be considered~\cite{AurellBavieraHammarlidServaVulpiani,AurellBavieraHammarlidServaVulpiani2}.

\bigskip
\noindent
{\bf Acknowledgment}

\noindent
We thank Angelo Vulpiani and Yi-Cheng Zhang
whose comments and suggestions have greatly contributed 
to realization of this work.
We also thank Roberto Baviera, Michele Pasquini and Frantis{\v e}k
Slanina for an infinite number of discussions concerning optimal strategies.

\end{document}